# Hydrodynamics calculations of clusters of drops using the finite volume method

Alejandro Acevedo-Malavé

Centro Multidisciplinario de Ciencias, Instituto Venezolano de Investigaciones Científicas (IVIC), Mérida 5101, Venezuela

**ABSTRACT**  In this paper is described a numerical scheme that is used to simulate the coalescence process between clusters of water drops immersed in a continuous hydrocarbon phase (n-heptane). Two different values for the initial velocity of the drops were chosen. Depending of the initial velocity of collision some scenarios emerge, such as: permanent coalescence, formation of satellite drops etc.  For some snap shots the streamlines are calculated for the different process of permanent coalescence. These streamlines allow the understanding of the dynamics of the droplets immersed on the n-heptane phase.  **KEY WORDS** Numerical simulation, finite volume method, Navier-Stokes, coalescence, droplets.

## INTRODUCTION

A molecular simulation study of the mechanism of droplets covered with a surfactant monolayer coalesce is reported[16]. The authors proposed a model system where the rate-limiting step in coalescence is the rupture of the surfactant film. For this numerical study one made use of the dissipative particle dynamics method using a coarse-grained description of the oil, water, and surfactant molecules. The authors found that the rupture rate is highest when the surfactant has a negative natural curvature, lowest when it has a zero natural curvature, and lying in between when it has a positive natural curvature. The coalescence of two condensing drops and the shape evolution of the coalesced drops is considered[8]. Image analyzing interferometry is used to study the coalescence of two drops of 2-propanol and the shape evolution after the coalescence is found to be driven by the capillary forces inside the drop. A method to study the dynamics of liquid drops by a numerical integration of the Navier-Stokes equations is proposed[7]. This author examined the motion of droplets with the application to the raindrop problem. The study was restricted to the collision of equal-sized drops along their line of centers. Numerical solutions were developed to study the rebound of water droplets in air. It is found that, except for a small viscous effect, the Weber number of the drops determines the dynamics of the collision and the bounce time. The formation of a liquid bridge during the coalescence of droplets is studied[4]. In this paper, the authors considered a mathematical model where the pressure singularity is removed at the instant of the impact for the coalescence of two viscous liquid volumes in an inviscid gas or in a vacuum environment. The formation of the liquid bridge is examined for two cases: (a) two infinitely long liquid cylinders, and (b) two coalescing spheres. In both cases the numerical solutions are calculated for the velocity and pressure fields, and the removal of the pressure singularity is confirmed. The Particle Image Velocimetry (PIV) experiments to study the coalescence of single drops through planar liquid/liquid interfaces was reported[13]. Sequences of velocity vector fields were obtained with a high speed video camera and the subsequent PIV analysis. Two ambient liquids with different viscosities but similar densities were examined. After rupture, the free edge of the thin film receded rapidly, allowing the drop fluid to sink into the bulk liquid below. Vorticity generated in the collapsing fluid developed into a vortex ring, straddling the upper drop surface. The inertia of the collapse deflected the interface downward before it rebounded upward. During this time, the vortex core split so that part of its initial vorticity moved inside the drop fluid while part of it remained in the ambient fluid above it. The velocity of the receding free edge was smaller for higher ambient viscosity and the pinching of the upper drop surface caused by the shrinking capillary ring wave was stronger when the ambient viscosity was lower. This resulted in a higher maximum collapse speed and higher vorticity values in the dominant vortex ring. An experimental investigation of binary collision of drops with emphasis on the transition between different regimes was proposed[15], which may be obtained as an outcome of the collision between droplets. In this study the authors analyze the results using photographic images, which show the evolution of the dynamics exhibited for different values of the Weber number. As a result of the experiment the authors reported five different regimes governing the collision between droplets are proposed: (i) coalescence after a small deformation, (ii) bouncing, (iii) coalescence after substantial deformation, (iv) coalescence followed by separation for head-on collisions, and (v) coalescence followed by separation for off-center collisions. An experimental study of the binary collision of water droplets for a wide range of Weber numbers and impact parameters was studied[2]. These authors identified two types of collisions leading to the drops separation, which can be reflexive separation and stretching separation. It is found that the reflexive separation occurs in head-on collisions, while stretching separation occurs in high values of the impact parameter. Experimentally, the authors reported the border between two types of separation, and also collisions that lead to coalescence. A model for drop coalescence in a turbulent flow field was reported[14] as a two-step process consisting in the formation of a doublet due to drop collisions followed by coalescence of the individual droplets occurring after the drainage of the intervening film by the action of van der Waals, electrostatic, and random turbulent forces. The turbulent flow field was assumed to be locally isotropic. A first-passage-time analysis was employed for the random process in the intervening continuous-phase film between the two drops. The first two moments of coalescence-time distribution of the doublet were evaluated. The average drop coalescence time of the doublet was dependent on the barrier due to the net repulsive force. The predicted average drop coalescence time was found to decrease whenever the ratio of the average turbulent force to repulsive force barrier became larger. The calculated coalescence-time distribution was broader with a higher standard deviation at lower energy dissipation rates, higher surface potentials, smaller drop sizes, and smaller size ratios of unequal drop pairs. A study on coalescence of unequal-size drops was reported[18]. In this study the coalescence of a drop with a flat liquid surface pinches off a satellite droplet from its top, whereas the coalescence of two equally sized drops does not appear to produce in this case a satellite drop. The authors found that the critical ratio grows monotonically with the Ohnesorge number and, it is reported, the experimental coalescence of two unequal-size droplets. A study about the coalescence of two equal-sized deformable drops in an axisymmetric flow using a boundary-integral method was carried out[17]. The thin film dynamics are simulated up to a film thickness of 10-4 times the non-deformed drop radius. The authors studied two different regimes for head-on collisions between the droplets. At lower capillary numbers the interfaces of the film between the drops remain in a circular flat form up to the film rupture. At higher capillary numbers the film becomes dimpled at an early stage of the collision process, also the rate of the film drainage slows down after the dimple formation. A study the coalescence collision of two liquid drops using a Galerkin finite element method in conjunction with the spine-flux method for the free surface tracking was reported[11]. The effects of some parameters like Reynolds number, impact velocity, drop size ratio, and internal circulation on the coalescence process are being investigated. The long time oscillations of the coalesced drops and the collision of unequal-size liquid drops are studied to illustrate the liquid mixing during the collision. A study about the collision of two mercury drops at very low kinetic energy using a fast digital and analog imaging techniques was carried out[12]. In this paper, the authors studied the time evolution of the surface shape as well as an amplified view of the contact region. The coalescence of two small bubbles or drops using a model for the dynamics of the thinning film in which both London-van der Waals and electrostatic double layer forces are considered[10,3]. A study on the effects of the direction of the applied electric field as well as the geometry of the electrodes was carried out[5]. In this study the angle between the electric field and the center line of two drops (θ), should be zero for the electrically induced force to attain its maximum attractive value. The maximum induced force is chosen large enough to deform the adjacent surfaces of the drop prior to coalescence. It has been shown experimentally and theoretically that the drop-drop attraction can occur also when θ is less than 54.7° or more than 125.3°. A method for determining the boundaries between the different outcomes that can be achieved in the collision of liquid drops setting the Weber number and varying the impact parameter is proposed[9]. A study about the behavior of a liquid-liquid interface and drop-interface coalescence under the influence of an electric field was carried out[6]. In this work the authors report experiments that show the measured electric current through water-sunflower oil and water-n-heptane systems, induced by an applied potential difference that increases linearly until a particular value, beyond which the measured current changes very rapidly. For various thicknesses of the water layer and the organic phase layer, the electric current corresponding to the turning points in the voltage-current characteristic curves, is between 3.7 and 18 nA. It is observed that the turning point of the voltage-current characteristic curve for a liquid-liquid system is caused by the formation of a cone at the interface. Above this turning point, the intensification of the local electric field above the tip of the cone is believed to be responsible for the very fast increase in the measured electric current. The measurement of the electric current can be used to monitor and control the behavior of a liquid-liquid interface, thus providing an optimum condition for instantaneous and single-staged drop-interface coalescence. The problem of the film drainage between two drops and vortex formation in thin liquid films was reported[1]. These authors also develop a simple model to describe dimple dynamics.

## Governing equations

The governing equations can be given by the continuity Eq. (1) And the momentum Eq. (2):

$$\frac{\partial}{\partial t}(r_\alpha \rho_\alpha) + \nabla \cdot (r_\alpha \rho_\alpha V_\alpha) = \sum_{\beta=1}^{N_p} \Gamma_{\alpha\beta}, \quad (1)$$

$$\frac{\partial}{\partial t}(r_\alpha \rho_\alpha V_\alpha) + \nabla \cdot (r_\alpha (\rho_\alpha V_\alpha \otimes V_\alpha))$$

$$= -r_\alpha \nabla p_\alpha + \nabla \cdot (r_\alpha \mu_\alpha (\nabla V_\alpha + (\nabla V_\alpha)^T)) + \sum_{\beta=1}^{N_p} (\Gamma^+_{\alpha\beta} V_\beta - \Gamma^+_{\beta\alpha} V_\alpha) + M_\alpha, \quad (2)$$

where $V$ is the velocity, $p$ is the pressure, $\mu$ is the viscosity, $\rho$ is the density, $r_\alpha$ is the volume fraction of the continuous phase $\alpha$, $N_p$ is the total number of phases, $M_\alpha$ describes the interphase forces acting on phase $\alpha$ due to the presence of other phases, $\Gamma^+_{\alpha\beta}$ represents the positive mass flow rate per unit volume from phase $\beta$ to phase $\alpha$. The unique term in the right hand of the equality on equation (1) occurs if the interphase mass transfer takes place.

CFX® (ANSYS® 15.0, ANSYS®, Inc. Southpointe 2600 ANSYS Drive Canonsburg, PA 15317 USA), a flow solver based on the finite volume method, was used to solve Eqs. (1) and (2). A rectangular mesh was used for this calculation with 360.000 elements. Inside the CFX module was defined the following constants: $\rho$(n-heptane)=683.8 Kg/m³, $\rho$(water)=1000.0 Kg/m³, $\mu$(n-heptane)=0.000408x10⁻³ Kg/m.s, $\mu$(water)=1.12x10⁻³ Pa.s, $r_{drop}$=15.0x10⁻⁶ m, $\sigma$(water/n-heptane)=49.1x10⁻³ N/m. Here $r_{drop}$ is the droplet radius, $\sigma$(water/n-heptane) is the interfacial tension of the system water-heptane.

## Coalescence and fragmentation of liquid drops

In order to model the collision of liquid drops some calculations were carried out using the Volume Finite Method. It was chosen the velocity of collision with values of 0.2 m/s which correspond to coalescence of water drops immersed in n-heptane continuous phase.

In Figure 1 is shown a sequence of snapshots for the collision between the water drops (blue color) immersed on a hydrocarbon continuous phase (white color). It can be seen that for a velocity of collision of 0.2 m/s the coalescence of drops is carried out, without the formation of the circular interfacial film that is reported in the literature. This behavior occurs because the liquid that drains out between the droplets has the time enough to drain and no fluid is trapped at the interface of the drops.

It can be seen at t=4.78e-7 sec the formation of two drops immersed inside the bigger drop of water, showing multiple oscillations due to the surface tension forces. At t=2.28e-6 sec the little drops recover its spherical form and there is no coalescence between these drops inside the bigger mass of water. After t=4.66e-5 sec the bigger mass of water takes a spherical form, and this form remains constant with the evolution of the dynamics.

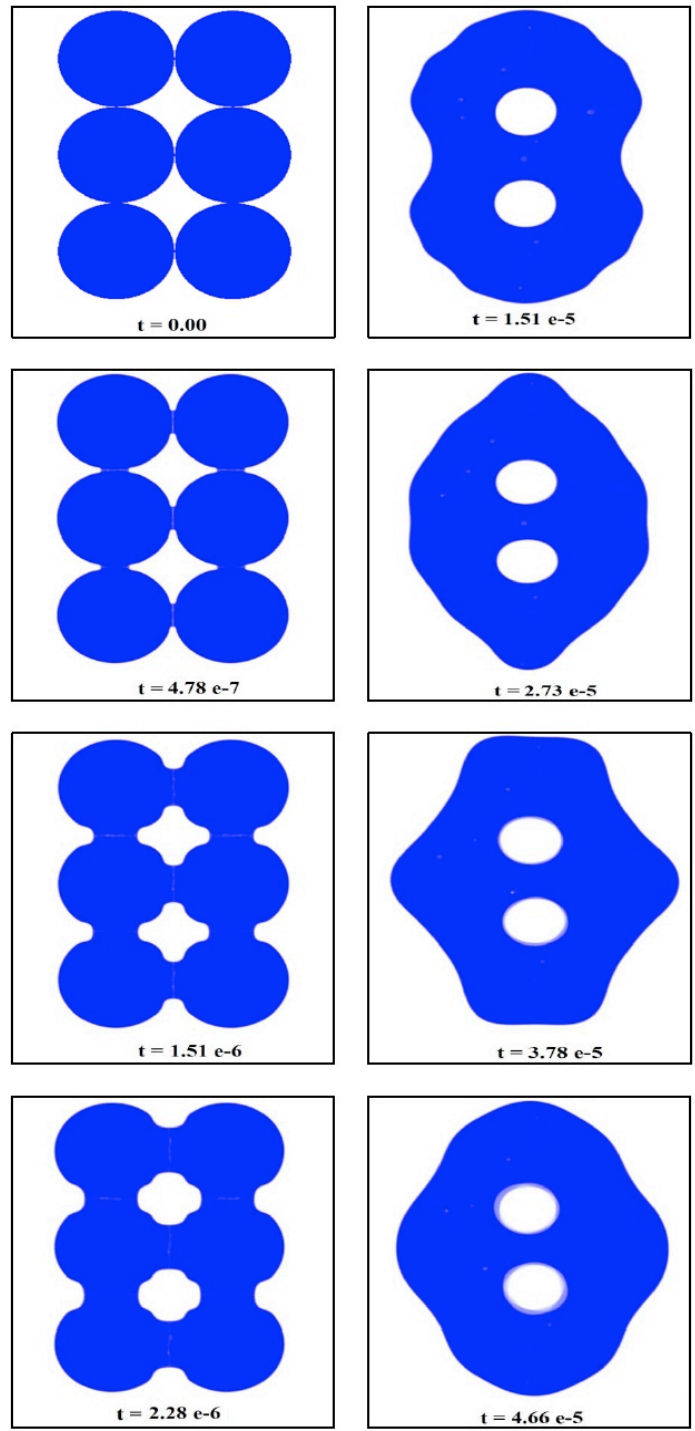

**Figure 1** Sequence of times showing the evolution of the collision between the drops with $V_{col}$ = 0.2 m/s. The time scale is given in seconds.

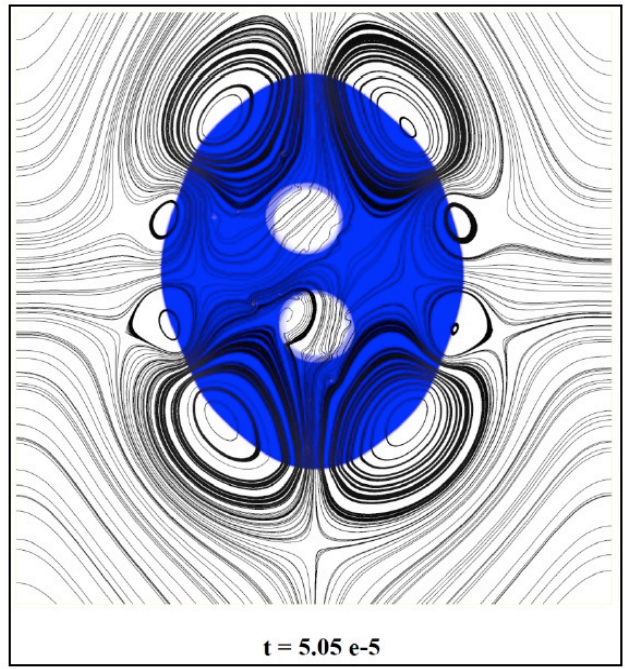

**Figure 2** Streamlines for the system water-heptane with $V_{col}$=0.2 m/s. The time scale is given in seconds.

In Figure 2 it can be seen the internal streamlines for the system of drops at t=5.05e-5 sec. Once the surfaces of the drops touch between them, the internal flux of the droplets has a circular form at many regions of the system. In fact, these forms of the flux are responsible for the curvature of the bigger drop in all directions of the system. At the end of the dynamics it can be seen that the bigger drop takes a perfect circular form.

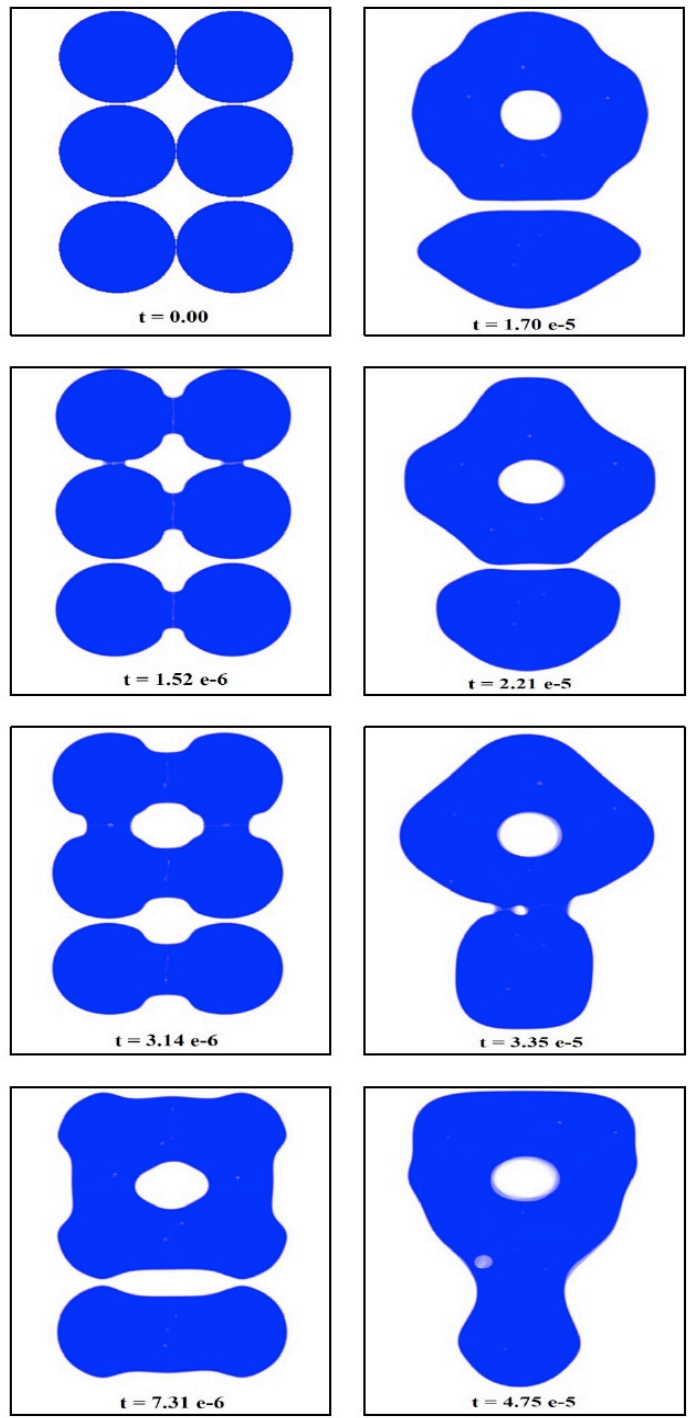

**Figure 3** Sequence of times showing the evolution of the collision between the drops with $V_{col} = 0.2$ m/s. The time scale is given in seconds.

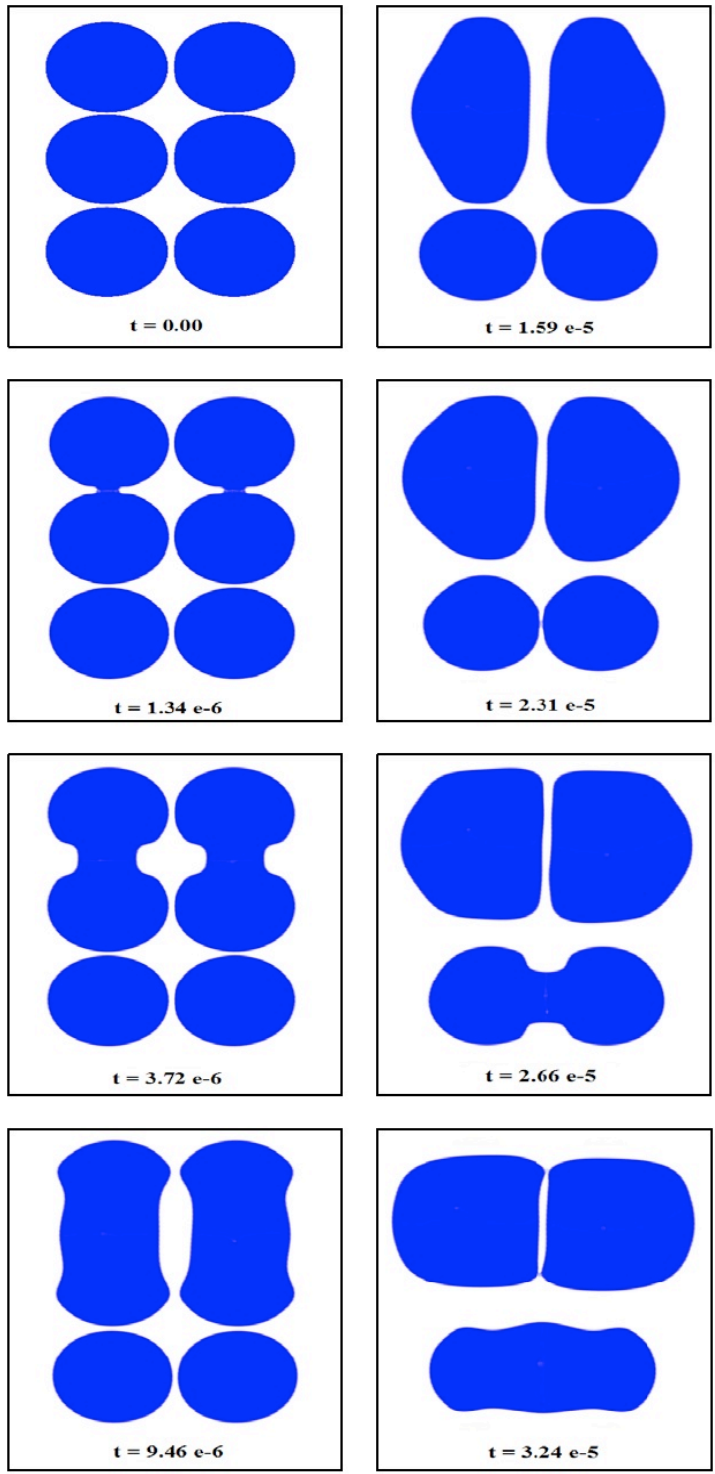

**Figure 4** Sequence of times showing the evolution of the collision between the drops with $V_{col} = 0.2$ m/s. The time scale is given in seconds.

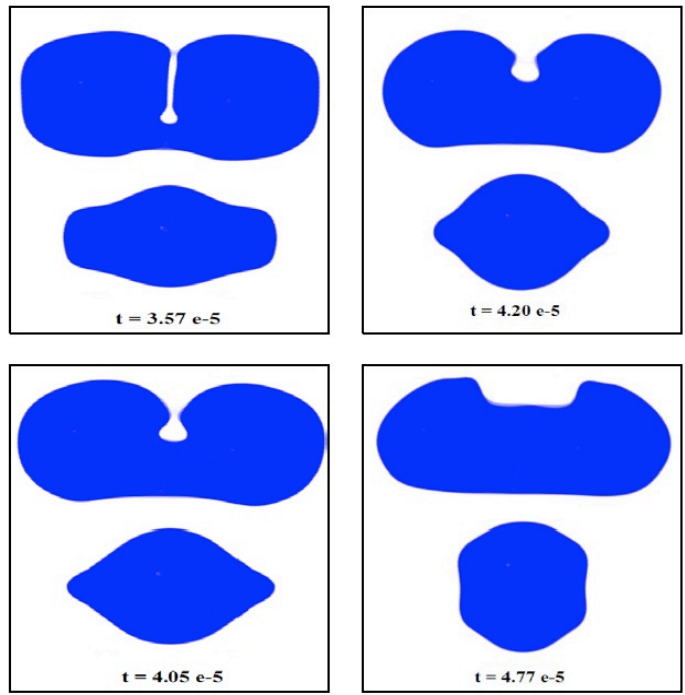

**Figure 5** Sequence of times showing the evolution of the collision between the drops with $V_{col}$ = 0.2 m/s. The time scale is given in seconds.

In Figure 3 is shown the evolution of the dynamics for the collision between the water drops immersed on the hydrocarbon continuous phase. It can be seen that for a velocity of collision of 0.2 m/s the coalescence of drops is carried out with the direction parallel to "x" axis. In this stage of the dynamics there is no the formation of the circular interfacial film that is reported in the literature. Equal that first case, this behavior occurs because the liquid that drains out between the droplets has the time enough to drain and no fluid is trapped at the interface of the drops. For this value of the velocity of collision the surface tension forces prevailing over inertial forces and the dynamics of the system water-heptane show some oscillations at the surface of the bigger drop.

In Figures 4 and 5 it can be seen the coalescence collision between two drops and after these, the bigger mass of water coalesce again between them, but at the end of the calculation these mass of water takes a spherical form and no coalescence is observed between them.

## CONCLUSIONS

In this paper a method for the study of hydrodynamical collisions of clusters of drops is presented. The equations of fluid dynamics were resolved numerically using the Finite Volume Method for a system composed of six water drops and the hydrocarbon (n-heptane) that represent a continuous phase. Depending on the initial conditions it was shown some different outcomes. The permanent coalescence with the occlusion of a drops of continuous phase inside the bigger drop of water and no coalescence was observed between the hydrocarbons droplets. The streamlines are shown for the flux inside and outside of the biggest drop and it can be seen that this flux can be divided in several regions where the flux is circular.

## REFERENCES

1. **Acevedo-Malavé, A., Sira, E., García-Sucre, M.** Film drainage between two drops: a vortex formation in thin liquid films. *Interciencia*, **34**:380-386, 2009.
2. **Ashgriz, N., Poo, JY.** Coalescence and separation of binary collisions of liquid drops. *J Fluid Mech*, **221**:183-190, 1990.
3. **Chen, JD.** A model of coalescence between two equal-sized spherical drops or bubbles. *J colloid Interface Sci*, **107**:209-220, 1985.
4. **Decent, SP., Sharpe, G., Shaw, AJ., Suckling, PM.** The formation of a liquid bridge during the coalescence of drops. *Int J Multi-phase Flow,* **32**:717-725, 2006.
5. **Eow, JS., Ghadiri, M.** The behaviour of a liquid–liquid interface and drop-interface coalescence under the influence of an electric field. *Coll Surf A: Physicochem Eng Aspects*, **215**:101-115, 2003.
6. **Eow, JS., Ghadiri, M.** Drop–drop coalescence in an electric field: the effects of applied electric field and electrode geometry. *Coll Surf A: Physicochem Eng Aspects*, **219**:253-265, 2003.
7. **Foote, GB.** The water drop rebound problem: Dynamics of collision. *J Atmos Sci*, **32**:390-403, 1974.
8. **Gokhale, SJ., Dasgupta, S., Plawsky, JL., Wayner, PC.** Reflectivity-based evaluation of the coalescence of two condensing drops and shape evolution of the coalesced drop. *Phys Rev E*, **70**:1-13, 2004.
9. **Gotaas, C., Havelka, P., Jakobsen, H., Svendsen, H.** Evaluation of the impact parameter in droplet-droplet collision experiments by the aliasing method, *Phys Fluids*, **19**:1-12, 2007.
10. **Li, D.** Coalescence between two small bubbles or drops. *J colloid Interface Sci*, **163**: 108-120, 1994.

11. **Mashayek, F., Ashgriz, N., Minkowycz, WJ., Shotorban, B.** Coalescence collision of liquid drops. *Int J Heat Mass Trans*, **46**:77-90, 2003.
12. **Menchaca-Rocha, A., Martinez-Davalos, A., Nuñez, R.** Coalescence of liquid drops by surface tension. *Phys Rev E*, **63**:1-14, 2001.
13. **Mohamed-Kassim, Z., Longmire, EK.** Drop coalescence through a liquid/liquid interface. *Phys Fluids*, **16**:1-12, 2004.
14. **Narsimhan, G.** Model for drop coalescence in a locally isotropic turbulent flow field. *J Coll Interf Sci*, **272**:197-210, 2004.
15. **Qian, J., Law, CK.** Regimes of coalescence and separation in droplet collision. *J Fluid Mech*, **331**:59-70, 1997.
16. **Rekvig, L., Frenkel, D.** Molecular simulations of droplet coalescence in oil/water/surfactant systems. *J Chem Phys*, **127**:1-14, 2007.
17. **Yoon, Y., Baldessari, F., Ceniceros, HD., Leal, LG.** Coalescence of two equal-sized deformable drops in an axisymmetric flow. *Phys Fluids*, **19**:1-12, 2007.
18. **Zhang, FH., Li, EQ., Thoroddsen, ST.** Satellite formation during coalescence of unequal size drops. *Phys Rev Lett*, **102**:1-11, 2009.

**Correspondencia:** Dr. Alejandro Acevedo-Malavé. Centro Multidisciplinario de Ciencias, Instituto Venezolano de Investigaciones Científicas (IVIC), Mérida 5101, Venezuela. Email: alaceved@ivic.gob.ve